\documentclass[12pt]{iopart}
\usepackage{graphicx}
\usepackage{dcolumn}
\usepackage{bm}
\usepackage{dsfont}
\usepackage{iopams}
\usepackage{pifont}
\usepackage{psfrag}
\usepackage{amssymb}

\newcommand\ket[1]{|{#1}\rangle}

\newcommand{\BE}{\begin{equation}}
\newcommand{\EE}{\end{equation}}

\newcommand{\skipc}[2]{}

\newcommand{\fig}[1]{figure~\ref{#1}}
\newcommand{\eq}[1]{(\ref{#1})}
\newcommand{\Sec}[1]{section~\ref{#1}}

\newcommand{\E}{\ensuremath{{\mkern1mu\mathrm{e}\mkern1mu}}}
\newcommand{\D}{\ensuremath{{\mkern1mu\mathrm{d}\mkern1mu}}}

\usepackage{color}

\begin{document}


\title{State selective detection of hyperfine qubits}

\author{Sabine W\"olk, Christian Piltz, Theeraphot Sriarunothai, and Christof Wunderlich}

\address{Department Physik, Naturwissenschaftlich-Technische Fakult\"at, Universit\"at Siegen, 57068 Siegen, Germany \\}%

\eads{sabine.woelk@physik.uni-siegen.de}

\date{\today}

\begin{abstract}
In order to faithfully detect the state of an individual two-state quantum system (qubit) realized using, for example, a trapped ion or atom, state selective scattering of resonance fluorescence is well established. The simplest way to read out this measurement and assign a state is the threshold method. The detection error can be decreased by using more advanced detection methods like the time-resolved method \cite{Myerson2008} or the $\pi$-pulse detection method \cite{Hemmerling2012}. These methods were introduced to qubits with a single possible state change during the measurement process. However, there exist many qubits like the hyperfine qubit of $^{171}Yb^+$ where several state change are possible. To decrease the detection error for such qubits, we develope generalizations of the time-resolved method and the $\pi$-pulse detection method for such qubits. We show the advantages of these generalized detection methods in numerical simulations and  experiments using the hyperfine qubit of $^{171}Yb^+$. The generalized detection methods developed here can be implemented in an efficient way such that experimental real time state discrimination with improved fidelity is possible.

\end{abstract}

\pacs{ 42.50.Dv, 37.10.Ty, 6.20.Dk }

\maketitle

\section{Introduction}

Quantum information processing can be divided into three steps: (i) the preparation of the system in a well defined state, (ii) the controlled time evolution of the system to carry out a desired algorithm, simulation or precision measurement, and, (iii) the readout of the quantum system. State selective detection is a  key ingredient for quantum information processing, not only necessary for readout, but also to verify the preparation of a system, to characterize the performance of quantum gates, or to perform an error correction algorithm. 
   
In ion traps state dependend scattering of resonance fluorescence is used for state selective detection. In this article we consider state selective detection of two internal ionic states labeled $\ket{d}$ and $\ket{b}$). Laser light drives a transition between one state of the qubit (the so called bright state) and a third fast decaying energy level. This leads to resonance fluorescence, if the ion was initially in the bright state. If the ion was initially in the other qubit state (dark state), no light, or only a small number of fluorescence photons is measured. This method for state selective detection can give rise to quantum jumps \cite{Dehmelt1986,Sauter1986,Bergquist1986}.

The simplest way to discriminate between the  bright and dark states of an  ion is the threshold method: if more than $n_c$ photons were registered during the measurement time $t_b$ we assume that the ion is bright, otherwise we assume it is dark. Due to the fact that a bright ion scatters photons only with a certain probability and dark states are not perfectly dark, due to background light not scattered by the ion and dark counts, statistical errors occur. This statistical error can be reduced by longer measurement times. However, the ion can change its state during the measurement which leads to additional systematic errors. These errors usually increase with longer measurement times.

In the context of this article we refer to a measurement of a qubit's state when resonant light is directed at the ion and an attempt is made to register fluorescence. The detection of the qubit state may, however, involve more than one measurement  and also additional coherent manipulations of the ionic internal states. When using the threshold method outlined above, the words ``measurement'' and ``detection'' have identical meaning. 

Several detection schemes were proposed and implemented to improve qubit detection by state selective resonance fluorescence. For example, Myerson et~al.  \cite{Myerson2008}  divided the total measurement time $t_b$ into several sub-bins of duration $t_s$ and calculated the probabilities $p_B$ ($p_D$) that  the measurement sequence is the result of an initially bright (dark) ion. A comparison of both probabilities reveals the more probable one, which determines the detection outcome.  We call this method the time-resolved detection method. It can also be applied to read out multi-qubits \cite{Burrell2010b}.

Another detection scheme was proposed and implemented by Hemmerling et~al \cite{Hemmerling2012} . They apply a $\pi$-pulse to the qubit states inverting their population after a first measurement followed then by a second measurement. Only results are kept with opposite results of the first and second measurement, all other results are discarded. Different methods to readout a single measurement can be used when applying this detection technique. However, here the threshold method seems to provide an advantage compared to the time-resolved method. In general, detection methods that discard doubtful results, such as this $\pi$-pulse method, have a higher failing probability (probability to get no answer or an incorrect answer) than detection methods that always provide an answer. Note that, by assigning randomly a state to measurement results  that were discarded, we obtain by chance some correct answers, which increases the overall probability to get a correct answer (success probability).
   Nevertheless, for some scenarios, the probability that the given answer is right is more important than the average probability of success. 
	
Both methods were designed for qubits  in the optical regime  such as $^{40}Ca^+$where the dark state can be transferred to the bright state via spontaneous decay, and the bright state is stable. As a consequence, only a single state change (from dark to bright) is possible. Therefore, the state of an initially bright state is fixed and the time dependent state of an initially dark ion can be described with a single parameter: the time $t$ at which the ion changes its state.

The present study was done in view of the widespread use of  hyperfine qubits (for example, $^9Be^+ $ \cite{Wineland2011},   $^{43}Ca^+$ \cite{Blatt2009,Allcock2013}, $^{137}Ba^+$ \cite{Dietrich2010},$^{171}Yb^+$ \cite{Balzer2006,Olmschenk2007,Ejtemaee2010,Meyer2012,Hensinger2013,Mount2013}) where the analysis of the measurement process is more complicated.

\begin{figure}
\begin{center}
 \includegraphics[width=0.4\textwidth]{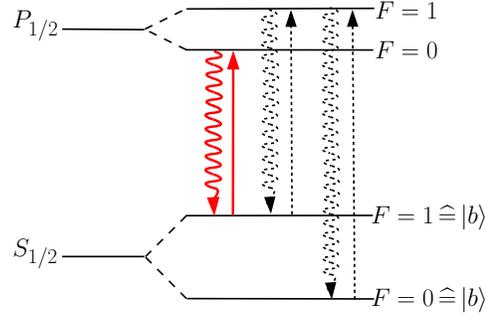}
\end{center}
 \caption{Level structure of $^{171}Yb^+$: levels $S_{1/2},F=0$ and one of the Zeeman states $S_{1/2},F=1$ form the hyperfine qubit. The transition from  $S_{1/2},F=1$  to  $P_{1/2},F=0$ (red arrow) is used for fluorescence detection and doppler cooling. Through off-resonant excitation to $P_{1/2},F=1$ and spontaneous decay (dashed arrows), the ion can change from the bright state $S_{1/2},F=1$ to the dark state $S_{1/2},F=0$ and vice versa. \label{fig:levels} }
\end{figure}

Hyperfine qubits  can change during the detection process from the bright state (for $^{171}Yb^+$ the state $S_{1/2},F=1$ as shown in \fig{fig:levels}) to the dark state ($S_{1/2},F=0$) and vice versa via off-resonant excitation and subsequent spontaneous decay. This leads to an increased number of parameters (times $t_j$ at which a state change takes place), due to the fact that not only one, but many state changes may occur.
Furthermore, when using time resolved measurements for detection, the photon-number distributions of individual time sub-bins are not independent of each other. As a consequence, the total probability of a measurement sequence is not given by the product of the probability distributions of the single sub bins. One way to deal with this problem is  to draw a decision tree and sum up the probabilities of all possible paths, which was done in \cite{Hemmerling2012}. However, in the case of several possible state changes, this leads to complicated formula: Assuming only a single possible state change per sub bin  for $M$ sub bins in total already leads to $2^M$ terms. As a consequence, we will not follow the calculation in Ref. \cite{Hemmerling2012} directly but we will use hidden Markov models instead, similar to Ref. \cite{Molmer2013}. In this way, the calculations can be performed in an efficient way.

In this paper, we generalize the ideas of Ref. \cite{Myerson2008} and Ref.\cite{Hemmerling2012} to two-level systems that allow several state changes during the measurement. Furthermore, we give an efficient expression to calculate the probability of a sequence of measurements  starting with the probability distributions of single measurements.  We apply this result to simulated measurement events and to experimental data obtained with trapped $^{171}Yb^+$ ions.

The paper is organized as follows: in \Sec{sec:time} we develop the mathematics necessary to generalize the time-resolved method to ions with several possible state changes during the measurement sequence. Then, we apply the detection scheme to a simulation of the hyperfine qubit of the $^{171}Yb^+$ ion followed by the description of the experimental realization of the improved time-resolved method  to trapped $^{171}Yb^+$ ions.  We finish  \Sec{sec:time} with a comparison of the improved time-resolved detection method developed in this paper with the original one. In \Sec{sec:pi} we generalize the $\pi$-pulse method in a similar way. Then we  apply the generalized $\pi$-pulse method to simulate detection of  the hyperfine qubit in $^{171}Yb^+$. We finish this section by comparing the generalized $\pi$-pulse method to a threshold method with two thresholds.


\section{Time-Resolved Detection \label{sec:time}}
Myerson et~al. consider in their work a qubit which can only change from the dark state to the bright state but not vice versa. Therefore, the probability, that the  sequence of measured photon numbers $\{n_k\}$ is the result of an initially bright state is given by $p_B(\{n_k\})=\prod P_B(n_k)$, where $P_B(n) $ is the probability distribution of measuring $n$ photons during a single sub bin. The probability, that the measurement sequence is the result of an initially dark ion is given by \cite{Myerson2008}
\BE
p_D=(1-\frac{t_b}{\tau})\prod\limits_{k=1}^M P_D(n_k)+ (\frac{t_s}{\tau})\sum\limits_{k=1}^M \prod\limits_{j=1}^{k-1} P_D(n_j)\prod\limits_{j=k}^{M} P_B(n_j).
\EE
Here, $P_D(n)$ is the photon distribution of the dark state, $M$ is the number of sub bins of duration $t_s$, $t_b$ is the total measurement time and $\tau$ the mean lifetime of the dark state (limited by spontaneous decay into the bright state). The term $(1-t_b/\tau)$ is the approximated  probability that the ion stays dark during the whole measurement and $(\frac{t_s}{\tau})$ is  the probability that the ion changes from dark to bright during a single sub bin of duration $t_s$.  As a consequence, the total probability is given by the sum of the probabilities of all possible paths (no state change, state change in first bin, $\cdots$). 

A straightforward way to generalize this formula to ions that allow several state changes would be to introduce a new summation index $j'$ for every possible state change. Assuming a maximal number $M$ of state changes, about $2^M$ terms needed to be summed up to calculate $p_D$ and $p_B$. This would lead to a fast increasing effort of data analysis and would make real-time readout very slow and adaptive schemes impossible for all practical purposes.

However, taking into account $2^M$ terms is not necessary, because the probability distribution of the $k^\textrm{th}$ sub-bin only depends on the state of the ion after the $(k-1)^\textrm{th}$ sub-bin and not on all previous sub-bins. Therefore, the probability distribution of the $k^\textrm{th}$ sub-bin can always be written as a sum of only two functions as we will show in this section.

\subsection{Generalization\label{sec:generalization}}
 
For our generalization, we assume that although the ion may perform several state changes during the total measurement time $t_b$, only a single state change may occur during a single sub-bin of duration $t_s$. This assumption can be justified by analyzing typical parameters relevant for the detection of $^{171}Yb^+$, or other ions  with hyperfine structure used for quantum information processing. In our experiments the mean life times of the states depend on the power of the laser beam used to scatter resonance fluorescence. Typically,  $\tau_B\approx 5.5\textrm{ms}$ for the bright state and $\tau_D\approx 50\textrm{ms}$ for the dark state. Using these lifetimes in our simulations, we  have found that from $10^5$ simulated bright (dark) ions $2\%$ ($0.2\%$) changed their state during a single sub-bin of duration $t_s=0.1$ms. None of them changed its state twice or more during a single sub-bin, which justifies our initial assumption. During  a total measurement time $t_b=3$ms, around $2\%$ of the ions change their state twice or more. 

Thus, the behavior of the ion during a single sub-bin is described by four probability distributions: (i) the probability of a bright ion staying bright
\BE
W_{BB}(t)=\E^{-t/\tau_B},\label{eq:WBB} \; t \in [0,t_s],
\EE
(ii) the probability $W_{BD}= 1-W_{BB}$ that a bright ion becomes dark, (iii) the probability of a dark ion staying dark
\BE
W_{DD}(t)=\E^{-t/\tau_D},\label{eq:WDD}\; t \in [0,t_s],
\EE
and (iv) the probability $W_{DB}=1-W_{DD}$ that a dark ion becomes bright.

Each of these four situations lead to different photon-number  distributions, which we determine as follows.
 The total measured photon rate of a dark ion  is the sum of the off-resonant fluorescence rate, the background scattering rate, and the dark count rate, and is given  by $R_D$. The total measured photon rate of a bright ion is given by $R_B+R_D$ (see App. \ref{RB}). If the ion does not change its state during  the measurement time $t_s$, then the probability of detecting $n$ photons is given by a Poisson distribution. For a bright ion we get
\BE
P_B(n)=\frac{[(R_B+R_D)\cdot t_s]^n}{n!}\E^{-(R_B+R_D)\cdot t_s}
\EE
and
\BE
P_D(n)=\frac{(R_D\cdot t_s)^n}{n!}\E^{-R_D\cdot t_s}
\EE
for a dark ion.

If the ion changes its state during $t_s$, then the probability of detecting $n$ photons is a superposition of Poisson distribution \cite{Burrell2010}
\BE
X(n)=\int\limits_{R_D\cdot t_s}^{(R_D+R_B)\cdot t_s} g(\lambda) \E^{-\lambda}\frac{\lambda^n}{n!}\D \lambda. 
\EE
For a bright ion becoming dark at exactly the time $t$, the mean photon number is given by
\BE
\lambda(t)=R_D\cdot t_s+R_B \cdot t.
\EE
Therefore, the weight function $g(\lambda)$ is given by
\begin{eqnarray}
g_{BD}(\lambda)&=&\frac{\D W_{BD}(t(\lambda))}{\D t}\left|\frac{\D t}{\D \lambda}\right|\\
&=&\exp\left[-\frac{\lambda-R_D\cdot t_s}{R_B\cdot\tau_B}\right]/(R_B\cdot \tau_B),
\end{eqnarray}
and we call the resulting function $X_{BD}(n)$.

Analogously, we obtain for a dark ion becoming bright 
\BE
g_{DB}(\lambda)=\exp\left[-\frac{(R_D+R_B) t_s-\lambda}{R_B\cdot\tau_D}\right]/(R_B \tau_D),
\EE
and we call the resulting function $X_{DB}(n)$. We note that the function $X(n)$ does not only contain information about the photon distribution but also about the probabilities of the ion to be bright or dark. As a consequence, the photon distribution of the $k^\textrm{th}$ sub-bin  is described by the matrix
\BE
O_k(n_k)=\left(\begin{array}{cc}W_{BB}P_B(n_k)& X_{DB}(n_k)\\ X_{BD}(n_k)& W_{DD}P_D(n_k)\end{array}\right).
\EE
These matrices have the property that the first (second) column contains information about ions that were bright (dark) before the measurement,  whereas the first (second) row contains information about ions that are bright (dark) after the measurement. 

This construction simplifies the calculation of the total probability $p_B(\{n_k\})$ ($p_D(\{n_k\})$) for the total series of measured photon numbers $\{n_k\}$ being the result of an initially bright (dark) ion to a simple matrix product as is shown below.

By defining
\begin{eqnarray}
  p_B(\{n_k\})&=&B_k^{(iB)}(\{n_k\})+D_k^{(iB)}(\{n_k\})\\
  p_D(\{n_k\})&=&B_k^{(iD)}(\{n_k\})+D_k^{(iD)}(\{n_k\})
 \end{eqnarray}
 where $B_k$ and $D_k$ stand for the probabilities of all possible paths where after the $k^\textrm{th}$ sub-bin the ion is in the bright or dark state, respectively, we find
\BE
\left(\begin{array}{cc}
 B_k^{(iB)}& B_k^{(iD)}\\ D_k^{(iB)}& D_k^{(iD)}\end{array}\right)=\prod\limits_{j=1}^k O_j(n_j).
\EE
This matrix product can be calculated very fast, if the function values of the four probability distributions ($W_{BB}P_B(n_k)$, $X_{DB}(n_k)$, $W_{DD}P_D(n_k)$, $X_{BD}(n_k)$) have been determined and stored for all possible photon numbers $0 \leq n_k\leq n_\textrm{max}$ in advance. In this way fast state detection on-the-fly is achievable which makes adaptive schemes possible, even in the presence of more than one state change.


\subsection{Simulation\label{sec:time_sim}}

In order to compare numerically different detection methods, we assume typical parameters for an $^{171}Yb^+$ ion: 
\BE
\tau_B=4.9\textrm{ms},\;\tau_D=56\textrm{ms},\;R_B=16/\textrm{ms}, R_D=0.3/\textrm{ms}, 
\EE
a sub-bin time of $t_s=0.1$ms, and a total measurement time of $t_b=3$ms or less.
The simulation of the detection process of a single initially dark or bright ion was performed in the following way: First, a random number generator randomly chooses the times $t_j$ at which the atom changes its state according to the probability distribution $W_{BD}$ or $W_{DB}$, respectively, until $\sum_j t_j>t_b$. In a second step, we generate the  photon numbers $\{n_k\}$  for each sub-bin measurement according to the Poisson-distribution  $P_B$ and $P_D$ for bright or dark ions. For sub-bins $k$ in which the ion changes from bright to dark, we use
\BE
\lambda_j = R_D\cdot ts+ R_B\cdot [t_j-(k-1)t_s]\textrm{ with } (k-1)ts < t_j < k\cdot t_s
\EE 
as mean photon number of the Poisson-distribution  and
\BE
\lambda_j = R_D\cdot ts+ R_B\cdot [(k\cdot t_s-t_j]\textrm{ with }(k-1)ts < t_j < k\cdot t_s
\EE
if the ion changes from dark to bright. In the last step, the different detection methods are applied to the generated data. A comparison of the initial state and the result of the detection methods determines the error. We define the error of bright ions by
\BE
\varepsilon_\textrm{bright}=\frac{\textrm{\#simulated bright ions detected as dark}}{\textrm{\# simulated bright ions}}.
\EE
The error of dark ions $\varepsilon_\textrm{dark}$ is defined analogously. 
 
\begin{figure}
\begin{center}
 \includegraphics[width=0.55\textwidth]{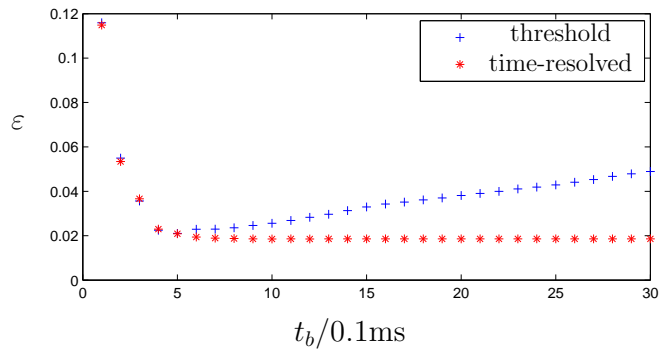}
\end{center}
 \caption{Comparison of the error of the threshold method (\textcolor{blue}{$+$}), and our improved time resolved method ($\textcolor{red}{*}$) for different total measurement times $t_b$ and constant sub-bin time $t_s=0.1$ms. \label{fig:tresh_time}}
\end{figure}
 
We simulated $10^5$ bright and $10^5$ dark ions and determined the average error $\varepsilon=(\varepsilon_\textrm{bright}+\varepsilon_\textrm{dark})/2$ of  the threshold method and our improved time-resolved method. For the threshold method we optimized the critical photon-number $n_c$ for each measurement time $t_b$.
As we can see in \fig{fig:tresh_time} the error is nearly equal for both methods for small measurement times. However, the minimum error of the threshold method with $\varepsilon_\textrm{thresh}\approx 2.1\%$ achieved for $t_b=0.8ms$ and $t_b=0.9ms$ is a little bit larger then the minimum error of the generalized time-resolved method with $\varepsilon_\textrm{time}\approx 1.85\%$  for $1ms\leq t_b \leq 3ms$. For long measurement times, $t_b$, the error of the threshold method increases whereas the error of the improved time-resolved method stays nearly the same.

This behavior can be explained in the following way: The threshold method assumes that there are no state changes and therefore, weights all measured photons in the same way, no matter when they arrive. However, in the limit of long times $t_b$, the photon distributions of an initially dark and an initially bright ion are indistinguishable  (see  \ref{appendix}), and therefore, the threshold method does not work anymore. In contrast, the time-resolved method takes state changes of the ion into account and puts more weight on early arriving photons.

\begin{figure}
\begin{center}
 \includegraphics[width=0.55\textwidth]{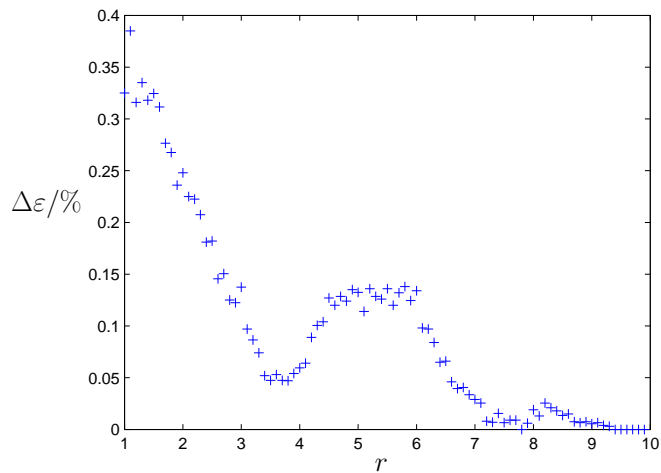}
\end{center}
 \caption{Behavior of the advantage $\Delta\varepsilon\equiv \varepsilon_\textrm{thresh}-\varepsilon_\textrm{time} $ of the generalized time-resolved method compared to the threshold method for different collection efficiencies $\eta=r\cdot \eta_0$. \label{fig:collefi}}
\end{figure}

The minimal detection error that can be reached depends on the experimental efficiency, $\eta$ with which scattered photons are collected. For numerically simulating the data shown in Fig. \ref{fig:tresh_time}, a collection efficiency $\eta_0=3.1\cdot 10^{-3}$ is used which was experimentally determined for the setup used in these investigations (see \Sec{sec:time_exp}). The collection efficiency includes the loss of fluorescence photons due to the limitation of the solid angle of the detector and its photo detection efficiency. It is interesting to see how an improved collection efficiency changes the detection error. For this purpose, we determine the behavior of the error  for different photon collection efficiencies $\eta=r\cdot \eta_0$  for the threshold method  and our time-resolved method where $r$ defines an enhancement factor. The two photon rate $R_B$ and  $R_D$ grow linearly with increasing collection efficiency.  Both detection methods benefit by the same magnitude from an enhanced collection efficiency. For $\eta=2\eta_0$ the error of both methods decreases by around $0.8\%$ to $\varepsilon_\textrm{thresh}=1.22\%$ and $\varepsilon_\textrm{time}=0.97\%$. An increasing collection efficiency leads to a   decreasing measurement time needed to distinguish between bright and dark ions. As a consequence, less state change will occur during the measurement time and    the advantage $\Delta\varepsilon\equiv \varepsilon_\textrm{thresh}-\varepsilon_\textrm{time}$ gained by the time-resolved method compared to the threshold method decreases (see \fig{fig:collefi}). Therefore, for r=9.9 both detection methods (generalized time-resolved and threshold) lead to nearly the same error of $\varepsilon \approx 0.33\%$. Here, we want to note, that the error for both methods are monotonical decreasing and the time-resolved method always better or as good as the threshold method. However, $\Delta\varepsilon$ is slightly oscillating due to the stepwise change of the threshold for different optimal measurement times $t_b$.

	
\subsection{Experimental Results\label{sec:time_exp}}

To determine the error rates of the threshold method and the generalized time-resolved method experimentally we capture a single $^{171}Yb^+$ ion in a Paul trap, laser cool it, prepare it in the dark or bright state, measure the number of  photons $ n_j$ arriving  during consecutive sub bins $j$ of duration $t_s=0.1ms$  and apply the different detection methods (for a detailed account of the experimental setup see \cite{Khromova2012}). Each measurement starts with the preparation of the ion in the dark state by driving the transitions $S_{1/2},F=1 \leftrightarrow P_{1/2},F=1$ (see \fig{fig:levels}) using laser light near 369.5 nm (preparation laser) and subsequent spontaneous decay of the ion to the dark ground state $S_{1/2},F=0$. For the preparation of a bright ion we us rapid adiabatic passage \cite{RAP} to transfer the state of the ion from dark to bright. Then the time dependent fluorescence on the resonance $S_{1/2},F=1 \leftrightarrow P_{1/2},F=0$ is measured using again laser light near 369.5 nm (measurement laser, detuned by 2.1 GHz relative to the preparation laser light). After each measurement we cool the ion with Doppler-cooling on the $S_{1/2},F=0 \leftrightarrow P_{1/2},F=0$ transition before preparing the next state.

To apply the time-resolved method to experiments, we have to determine not only the photon rates for the bright and the dark ion, but also the rate at which state changes occur from bright to dark and vice versa. All rates depend on the intensity of the measurement laser. To achieve this task we first measure the time dependent fluorescence for initially dark and bright ions for a total time of 10 ms, divided into 30 sub-bins of duration $t_b=1/3$ms. The average photon number per sub-bin is given by
\BE
\overline{n}_B(t)= a+b \E^{-t/\tau}
\EE
for an initially bright ion and 
\BE
\overline{n}_D(t)= a-c \E^{-t/\tau}  
\EE
for an initially dark ion. The parameters $a$ and $\tau$ are the same in both cases. Here we note, that in the presence of the fluorescence laser all states turn into a steady state  in the long time limit.  In this state,  on average the same number of ions turn from dark to bright  as turn from bright to dark. As a consequence the photon rate  in the long time limit does not depend on the initial state and is determined by $a$. The mean lifetime $\tau=\tau_B\tau_D/(\tau_B+\tau_D)$ determines the time scale to reach this steady state.
With the help of 
\BE
A\equiv \frac{b/c}{1+b/c} \quad\textrm{and}\quad B\equiv 1-A
\EE
we can estimate the lifetime  $\tau_B=\tau/A$ of the bright state and $\tau_D=\tau/B$ of the dark state (for derivation see \ref{appendix}). The coefficients $A,B$ determine the probability to be in the bright or dark state for the steady state.

\begin{figure}
\begin{center}
\includegraphics[width=0.55\textwidth]{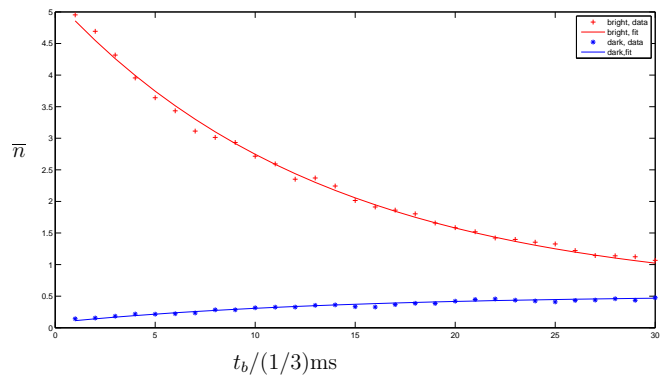}
\end{center}
\caption{Long time measurement of the time dependent fluorescence  of an average bright ion (\textcolor{red}{+}) and an average dark ion (\textcolor{blue}{$\ast$}) and their simultaneous fit ( lines) for a laser power of $36\mu$W and a beam diameter given by $174\mu$m (measurement laser near 369.5 nm). With a collection efficiency of $\eta_0=3.1 \cdot 10^{-3}$  the average measured photon scattering rates are $R_B=16/$ms for the bright state and  $R_D=0.3/$ms for the dark state (see \Sec{sec:generalization} for the definition of the scattering rates). }
\label{fig:doppelfit_lang}
\end{figure}

We measured the time dependent fluorescence for  a single ion $2000$ times consecutively prepared in the bright state or $2000$ times in the dark state to determine the time-dependent mean photon rates.  The measured average photon number per sub-bin is shown in \fig{fig:doppelfit_lang}. We have fitted $\overline{n}_B$ and $\overline{n}_D$ simultaneously, which means we minimized 
\BE
\sqrt{\sum\limits_j (\overline{n}_{B,j}-\overline{n}_B(j\cdot t_S))^2+\sum\limits_j (\overline{n}_{D,j}-\overline{n}_D(j\cdot t_S))^2}.
\EE
Here, $\overline{n}_{B,j},\overline{n}_{D,j} $ are the measured mean photon number of sub bin $j$ for an initially bright or dark ion, respectively. 

For a laser power of 36$\mu$W focused to a beam diameter of 174$\mu$m (measurement laser) the fit leads to the parameter $a=0.515$, $b=4.68$, $c=0.434$, $\tau=4.50$ms. As a consequence, the lifetimes of the dark and bright ion in our experiment are given by $\tau_B=4.92$ms and $\tau_D=53.1$ms. With these parameters, we are now able to apply the time-resolved detection method.
For this purpose, we  measure a total of $9\times10^3$  bright ions and  $9\times10^3$  dark ions, always a single bright and a single dark ion in turns. At the measurement laser intensity quoted above  and with a collection efficiency of $\eta_0=3.1\cdot 10^{-3}$  the average measured photon rates are $R_B=16/$ms and  $R_D=0.3/$ms. 

\begin{figure}
\begin{center}
\includegraphics[width=0.55\textwidth]{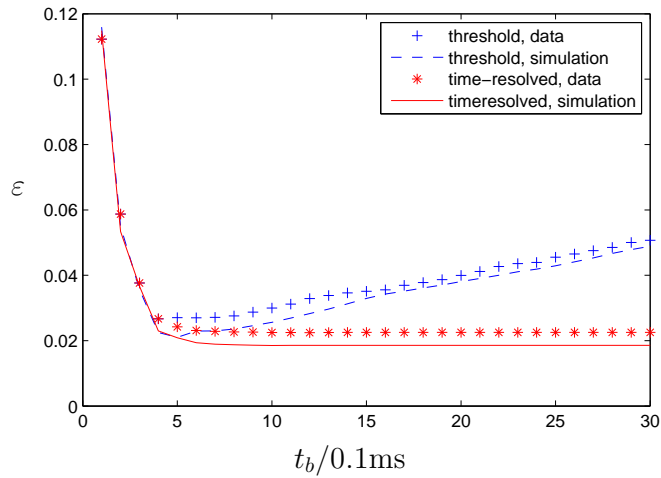}
\end{center}
\caption{Experimental errors of the threshold (\textcolor{blue}{+})and the time-resolved method  (\textcolor{red}{$\ast$}) in comparison to each other and to simulations}
\label{fig:auswertung}
\end{figure}

We evaluate the data with both detection methods and estimate the error depending on the number of measurement bins used for the detection methods. In \fig{fig:auswertung} we see that the experimental results show the same qualitative behavior as the simulations. These simulations were done for both methods, using the  experimental parameters given in the last paragraph. The minimal experimental error of the improved time resolved method is determined as $\varepsilon_{time}=2.24\%$, and, thus smaller than the error of the threshold method given by $\varepsilon_{thres}=2.67\%$.  The simulations of both detection methods reach smaller errors due to the fact that the simulation does not consider preparation errors or fluctuations of laser power or frequency.

In Ref.\cite{Ejtemaee2010}  Ejtemaee et~al. report how they optimized the laser intensity to get the best detection efficiency. Optimal detection was achieved for a fluorescence rate of $R_B\approx 25/ms$.  The collection efficiency in \cite{Olmschenk2007} was given by $2.9\cdot 10^{-3}$ and was therefore approximately equal to the experiment reported here. This means that the intensity of the measurement laser differed. Although our  experimental parameter seem to differ slightly from the optimal one, we achieve detection efficiencies exceeding 97$\%$ similar to \cite{Ejtemaee2010}. Therefore, by optimizing the experimental parameter and using the general time-resolved detection method, it should be possible to exceed the detection efficiency of $97.9\%$ measured by Olmschenk et~al. \cite{Olmschenk2007}.

Recently, Noek et~al. were able to improve the state detection efficiency of hyperfine qubits with $^{171}Yb^+$ dramatically to $\varepsilon= 0.085\%$ \cite{Noek2013}. The main reason of this improvement is based on an improved photon collection efficiency $\eta$ which is around 10 times larger than the ones of our experiment or the experiment done by Ejtemaee et~al. and the reduction of background photons. The reduction of the error $\varepsilon$ with our generalized time-resolved detection method is small compared to the reduction gained by a higher photon collection efficiency. Nevertheless, our measurement scheme is very useful, since for every fixed collection efficiency, it is still able to reduce the error over a wide range of $\eta$ as shown in \fig{fig:collefi}. As a consequence, even if an improvement of the collection efficiency is not possible due to structurally engineered reason, the detection error can be reduced by using our generalized time-resolved method.

\subsection{Comparison with original algorithm}

In the previous sections we showed how to generalize the time-resolved detection method from Ref. \cite{Myerson2008} and applied it to simulations and to experimental data. Our experiments showed (see \Sec{sec:generalization}), that only a small fraction of the ions execute several state changes. Therefore,  we compare now the generalized method with the original method to investigate whether additional effort of the generalized time-resolved method leads to better results of our experiment.

\begin{figure}
\begin{center}
\includegraphics[width=0.55\textwidth]{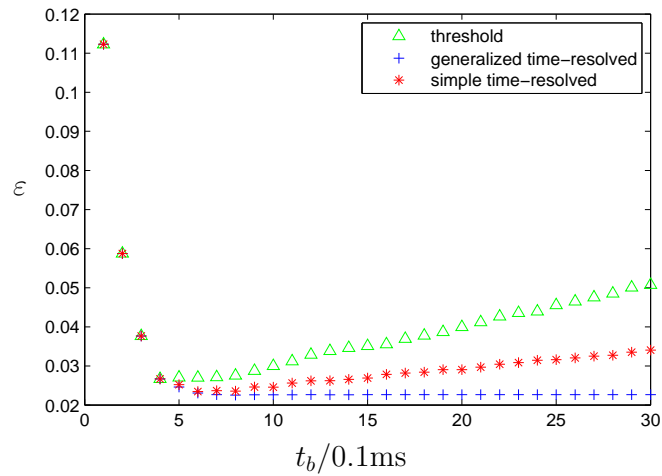}
\end{center}
\caption{ Comparison of the experimental data evaluated with the original  time-resolved method (\textcolor{red}{*}) that considers only one possible state change from bright to dark, our generalized time-resolved method (\textcolor{blue}{+}) that  considers several possible state changes from bright to dark and vice versa, and the threshold method (\textcolor{green}{$\triangle$})}
\label{fig:simple_general}
\end{figure}

In \fig{fig:simple_general} we compare the evaluation of our experimental data (see \Sec{sec:time_exp})  using different detection methods. The simple time-resolved method gives better results than the threshold method, and the generalized time-resolved method leads to a further reduction of the detection error. In particular, for larger measurement times $t_b$, the error of the simple time-resolved methods starts to increase significantly, whereas the error of the generalized method stays nearly the same, making the later more robust.  The minimal error achievable with the simple time-resolved method with min($\varepsilon_{simple})=2.34\%$ is slightly larger than the error of the generalized method with min($\varepsilon_{general})=2.24\%$. 

In order to investigate the significance of this difference observed in the experimental data, we performed 20 simulations with $10^5$ ions each with the parameters determined in \Sec{sec:time_exp} and evaluated them with the simple and the generalized time-resolved method. For the simple time-resolved method we found an average error of $\overline{\varepsilon}_{simple}=1.92\%$ with a variance of $\Delta \varepsilon_{simple}=0.026\%$. For the generalized time-resolved method we found $\overline{\varepsilon}_{simple}=1.80\%$ with a variance $\Delta \varepsilon_{simple}=0.029\%$. Again, our simulations does not take into account preparation errors or errors due to the drift  of  the laser frequency and therefore lead to smaller errors than our experiment.

In summary, we see in \fig{fig:auswertung} and \fig{fig:simple_general} that  we benefit more and more from the  time-resolved method (the simple one and especially our generalized one) compared to the threshold method when the optimal time $t_b^\textrm{opt}$ necessary to collect enough photons for state discrimination increases compared to lifetime $\tau$ of the state. For $t_b^\textrm{opt}\ll \tau$  no difference between the three detection methods (threshold, simple time-resolved, generalized time-resolved) exists. For an increasing $t_b^\textrm{opt}$ a benefit from the time-resolved method becomes visible. For even larger $t_b^\textrm{opt}$ the difference between the simple and the generalized time-resolved method becomes visible. Since $t_b^\textrm{opt}$ depends on the ratio between the fluorescence rate $R_B$ and the dark count rate $R_D$, it is also this ratio together with the fluorescence rate $R_B$  which decides if the time-resolved methods is advantageous. 

Increased laser power leads to a decrease of $t_b^\textrm{opt}$  as well as of $\tau$, and therefore to a faster measurement. However, this also changes the minimal error achievable. A higher fluorescence rate does not, in general, lead to a smaller detection error \cite{Ejtemaee2010}. 

An increased collection efficiency $\eta$ leads also to a shorter measurement time, and also to a decreasing minimal error. Since the lifetime $\tau$ is independent of $\eta$, an increased $\eta$ may also decrease the advantage of the time-resolved method (see \fig{fig:collefi}).

\section{$\Pi$-Pulse Detection \label{sec:pi}}

Another way to increase the detection efficiency is to perform a detection followed by a $\pi$-pulse and a second detection as described in \cite{Hemmerling2012}. Only results with different detection outcomes for detection one and two are considered, detections with the same outcome for both detections are excluded.
Similarly to the previous section, we have to generalize the results of \cite{Hemmerling2012} to an ion that cannot only change from the dark state to the bright state but also vice versa in order to consider this method for ions where several state changes during the detection process are possible. 


\subsection{Generalization}
The calculations of the error  in \cite{Hemmerling2012} where done with the help of a decision tree. However, the possibility of several state changes (instead of a single state change) increases the number of possible branches exponentially. In the presence of several possible state changes, it is, therefore, not useful to draw a decision tree and sum up all possibilities. Instead, we develop a more efficient way to determine the error with the help of hidden Markov models as we show in this section. 

We start to describe the measurements with the help of matrices. The important variables for the $\pi$-pulse detection scheme are: (i) the initial state, (ii) the state after the measurement, (iii) the detection outcome. Therefore, the probability of detecting a bright ion is in general a sum of $4$ probabilities: (i) an initially bright ion stays bright and is correctly  detected as bright ($R_{BB}$), (ii) an initially bright ion turns dark and is correctly  detected as bright ($R_{BD}$), (iii) an initially dark ion stays dark and is falsely detected as bright ($F_{DD}$),(iv) an initially dark ion turns bright and is falsely detected as bright ($F_{DB}$). These four probabilities form the matrix $M_B$ given by
\BE
M_B\equiv \left(\begin{array}{cc} R_{BB}& F_{DB}\\ R_{BD} & F_{DD}                 
                \end{array}
\right),
\EE
which does not only help us to calculate the probability of detecting a bright ion but also contains the information about the state of the ion after the measurement. By describing a bright ion  by the vector $v_B\equiv (1,0)^T$ and a dark ion by $v_D\equiv (0,1)^T$ the probability that an initially bright ion is detected as bright is given by $p_{BB}=p^{(1)}+p^{(2)}$ with
\BE
\left(\begin{array}{c}p^{(1)}\\p^{(2)}\end{array}\right)\equiv \left(\begin{array}{cc} R_{BB}& F_{DB}\\ R_{BD} & F_{DD}    \end{array}\right)\left(\begin{array}{c}1\\0\end{array}\right)=M_B v_B,
\EE
where $p^{(1)}$ is the probability that the ion is in the bright state after the measurement and detected as bright and $p^{(2)}$ is the probability that the ion is in the dark state after the measurement and detected as bright.

Analogously, the probability of detecting a dark ion and its state after the measurement is determined by the matrix
\BE
M_D\equiv \left(\begin{array}{cc} F_{BB}& R_{DB}\\ F_{BD} & R_{DD}                 
                \end{array}
\right).
\EE
The $\pi$-pulse that turns dark states into bright states and vice versa is described by the matrix
\BE
M_{\pi}\equiv \left(\begin{array}{cc} \epsilon_\pi & 1-\epsilon_\pi\\ 1-\epsilon_\pi & \epsilon_\pi                     
                    \end{array}
\right)
\EE
where $\epsilon_\pi$ is the error of the $\pi$-pulse. 

The process that an initially bright ion is falsely detected as dark by the $\pi$- pulse method is therefore described by the vector
\BE
f_B=M_B M_\pi M_D v_B\label{def:fB},
\EE
where the total error is given by the sum of the entries of $f_B$.
In the same way, the probability of detecting the bright ion correctly is determined by
\BE
r_B=M_D M_\pi M_B v_B\label{def:rB},.
\EE
In addition to the ions that are correctly detected and the ions with a wrong detection result, there exists a third category: ions that are ignored, because the detection result of the first and second detection are equal. We call the ions that are not ignored the remaining ions.

When determining the error associated with a scheme where some ions are ignored (i.e. an inconclusive result is obtained for these ions), the ratio between the number of  wrong answers and the total number of detected ions (which is the sum of remaining and ignored ions) may not be a useful criterion. To illustrate this point, we consider a simple  example:  a  possible worst case scenario is that the detection gives a wrong answer or no answer at all, but never the correct answer. In this case, despite the fact that one never obtains a correct answer, the detection error 
could be found to be small.
Therefore, in what follows we consider instead the relative error defined as the ratio between the number of wrong results and the number of remaining ions.  For this purpose, we calculate how many detections of the remaining data lead to a wrong result. This relative error determines how reliable the result of the detection is, if we get one. This relative error  is given by
\BE
\epsilon_B^{\textrm{rel}}=\frac{f_B^{(1)}+f_B^{(2)}}{f_B^{(1)}+f_B^{(2)}+r_B^{(1)}+r_B^{(2)}},
\EE
where $f_B^{(j)}$ and $r_B^{(j)}$ denotes the $j^\textrm{th}$ entry of the vector $f_B$  \eq{def:fB} and $r_B$ \eq{def:rB}, respectively.
Analogously, we calculate the relative error for detecting a dark ion as follows:
\begin{eqnarray}
 f_D&=&M_D M_\pi M_B v_D\\
 r_D&=&M_B M_\pi M_D v_D\\
 \epsilon_D^{\textrm{rel}}&=&\frac{f_D^{(1)}+f_D^{(2)}}{f_D^{(1)}+f_D^{(2)}+r_D^{(1)}+r_D^{(2)}}
\end{eqnarray}
and the total error $\epsilon^{\textrm{rel}}=(\epsilon_B^{\textrm{rel}}+\epsilon_D^{\textrm{rel}})/2$.

\subsection{Simulation}

For single detections before and after the $\pi$-pulse we use either the threshold or the generalized time-resolved method. We simulate again $10^5$ bright and $10^5$ dark ions to determine the matrices $M_B$ and $M_D$ for different measurements times $t_b$ and fixed sub-bin time $t_s=(0.1/3)$ms and optimized $n_c$. We assume an error of $\varepsilon_\pi=0.02$ \cite{Hemmerling2012} for the $\pi$-pulse. With the help of these matrices we are able to determine the overall error $\varepsilon$.  

\begin{figure}
\begin{center}
\includegraphics[width=0.65\textwidth]{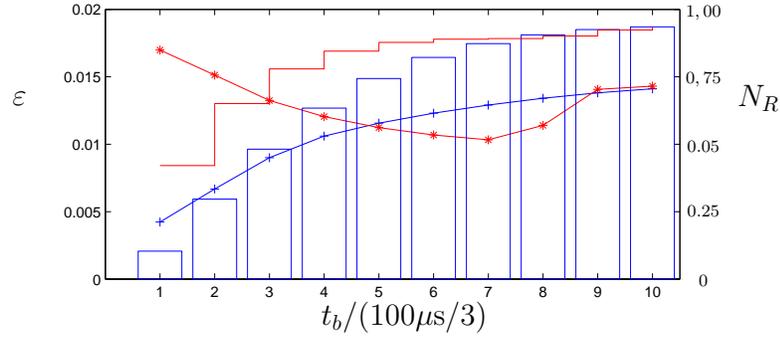}
\end{center}
\caption{Comparison of the relative error  (left scale ) of the generalized time-resolved method  combined with the $\pi$-pulse method (\textcolor{red}{$*$}) and the threshold method combined with the $\pi$-pulse method (\textcolor{blue}{$+$}), for different total measurement times $t_b$ and fixed sub-bin time $t_s=0.1/3$ms. The detection efficiency $N_R$ for the generalized time-resolved (red stair diagram) and the threshold method (blue bar diagram)  correspond to the right scale\label{fig:pi_time_tresh}}
\end{figure}

Similar to Ref.\cite{Hemmerling2012}, we find that the $\pi$-pulse method can reduce the relative error  of the threshold method as well as of the generalized time-resolved method. For the generalized time-resolved method we get, for the above mentioned parameter set, a minimal error of $\varepsilon=1.0\%$ compared to $1.85\%$ (see \Sec{sec:time_sim}) without the $\pi$-pulse. As we can see in \fig{fig:pi_time_tresh} the error of the $\pi$-pulse method combined with the improved time-resolved method increases for smaller time scales whereas the combination of $\pi$-pulse method and the threshold method seems to decrease. The minimal error  $\varepsilon=0.4\%$ displayed in \fig{fig:pi_time_tresh} is obtained using the threshold $n_c=1$. 

It is also important to compare detection efficiency  defined as 
\BE
N_R=\frac{\# \textrm{remaining ions}}{\#\textrm{total  ions}}
\EE
for the generalized time-resolved method (red stair diagram) and the threshold method (blue bar diagram) also displayed in \fig{fig:pi_time_tresh} and corresponding to the right scale. For small times $t_b$ only a few ions are remaining if the threshold method is used, e.g. only $10\%$ are remaining for the minimal error achieved for $t_b=100\mu s/3$.  This means that a small relative error is obtained at the expense of the number of ions that yield a conclusive detection result. If the generalized time-resolved method is used, also for small times more than $40\%$ of the ion lead to a conclusive measurement result. For large measurement times, $N_R$ is nearly equal for both methods.

\begin{figure}
\begin{center}
\includegraphics[width=0.65\textwidth]{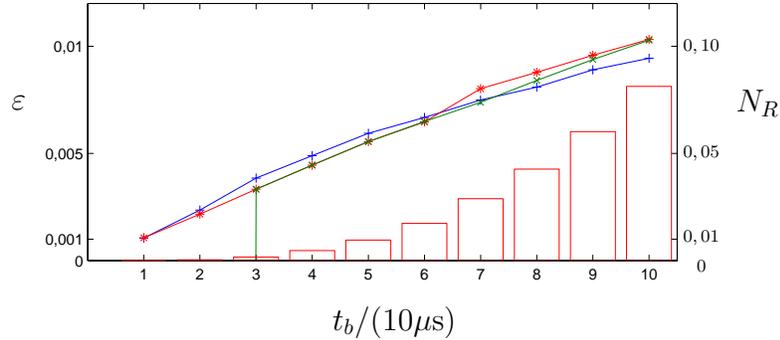}
\end{center}
\caption{Relative error (left scale) of the  the threshold method combined with the $\pi$-pulse method for different total measurement times $t_b$ and different thresholds $n_c=1$ (\textcolor{blue}{$+$}), $n_c=3$ (\textcolor{red}{$*$}),  $n_c=5$ (\textcolor{green}{$\times$}). The red bar diagram displays the detection efficiency $N_R$ for $n_c=3$ (right scale). \label{fig:pi_tresh}}.
\end{figure}

The problem of small numbers of remaining ions  is further illustrated by considering the relative error for even smaller measurement times. 
As dispaleyed in \fig{fig:pi_tresh}, the error of the $\pi$-pulse method using the threshold method decreases for even smaller time-scales independently of the chosen threshold $n_c$. However, we have to be  careful with this statement because for very small total measurement times $t_b$ it becomes nearly impossible to detect bright ions. As a consequence, the relative error is not well defined anymore. For example, in \fig{fig:pi_tresh} no error for $nc=5$ and $t_B< 30\mu$s is displayed, because all data has been neglected. However, in these cases, we get no information about the state of the ion and therefore it is also not useful to calculate the error. Even if there is some remaining data, we have to be careful: for example for $nc=1$ and $t_B=10\mu$s it was possible to calculate an error, but the remaining data was small: Out of $10^5$ bright ions, only $23$ were detected as bright in a single detection, and therefore only $0.023\%$ of the data was remaining. This means that (i) there exists a large statistical error in the calculation of the error and  (ii) we need many measurements before we get a statement about the state of the ion. 

In summary, the $\pi$-pulse method \cite{Hemmerling2012} has been generalized from qubit states where only  a single state change occurs during a measurement to states with several possible state changes. This  generalization is achieved by using methods from hidden Markov Models instead of decision trees. We find that  the  generalized $\pi$-pulse method can reduce the relative error of the threshold method as well as of the improved time-resolved method. 

\subsection{Double-threshold method\label{sec:double_tresh}}

Applying the $\pi$-pulse method to an ion like  $^{171}Yb^+$ is not straight forward, because the bright state is split into  three states $m_F=-1,0,+1 $ (see \fig{aufspaltung}).

Nevertheless, the general idea of dividing the measurement results not only into bright and dark states, but also into the group ``inconclusive result''  decreases the detection error (at the expense of the detection efficiency), which we show in this section by considering the double-threshold method.

For the double-threshold method we define two thresholds:(i) if the measured photon number $n$ obeys  $n\leq n_D$ we assume that the ion is dark, (ii) if $n>n_B$ we assume that the ion is bright. If $n$ is in between, we make no statement about the state of the ion and ignore this datum. 

\begin{figure}
\begin{center}
 \includegraphics[width=0.65\textwidth]{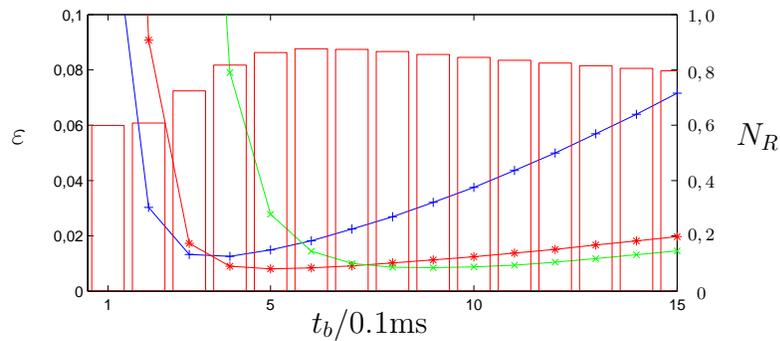}
\end{center}
 \caption{Error of the double threshold method for lower limit $n_D=0$ and different upper limits $n_B=1$ (\textcolor{blue}{$\times$}), $n_B=4$ (\textcolor{red}{$+$}), $n_B=10$ (\textcolor{green}{$*$})   for different measurement times $t_b$. The red bar diagram displays the detection efficiency $N_R$ for $n_c=4$ (right scale). \label{fig:tresh2l}}
\end{figure}

In \fig{fig:tresh2l} we display the result of simulations for a lower threshold $n_D=0$\footnote{our simulations showed that $n_D=0$ is the optimal lower threshold}. We first calculate the relative error $\varepsilon^\textrm{rel}= (\textrm{\# false results})/(\textrm{\# remaining results})$ for bright and dark ions before we average them to the total error
\BE
\varepsilon=\frac{\varepsilon^\textrm{rel}_B+\varepsilon^\textrm{rel}_D}{2}.
\EE
The error for a fixed threshold $n_B$ first rapidly decreases as a function of time, before it increases again slowly; for each fixed threshold $n_B$, there exists a minimum error. For a larger upper threshold $n_B$ we get a larger optimal time. The minimal error varies with the threshold $n_B$. It first decreases with increasing threshold $n_B$ and reaches a minimum before it increases again. Therefore, we have to optimize the threshold $n_B$ and the measurement time $t_b$ to get the minimum error similar to the normal threshold method.
For $\tau_B=4.9$ms, $\tau_D=56$ms, $R_B=16/$ms and $R_D=0.3/$ms we get a minimal error of $0.81\%$ for $n_B=4$ and $t_b=0.5$ms  with a detection efficiency of $N_R=0.86$. If we demand a detection efficiency of $N_R>0.8$ the minimal error achievable with the $\pi$-pulse method is only $\varepsilon=1.23\%$ which is worse than the error of the double threshold method. This may be caused be the additional error caused by the error of the $\pi$-pulse and the higher rate of state changes during two consecutive measurements, each of duration $t_b$, compared to a single measurement of duration $t_b$. However, the minimal error achievable with the double threshold method is limited, whereas the $\pi$-pulse method can reached arbitrary small errors at the cost of a decreasing detection efficiency. To beat the minimal error of the double threshold method with the $\pi$-pulse method, we have to tolerate a detection efficiency of $N_R<0.4$.

Both, the $\pi$-pulse method and the double-threshold method ignore some data to decrease the detection error, yet, they behave quite differently: For very small time scales, the $\pi$-pulse method neglects all data because it is not possible to detect bright states. The double-threshold method detects all dark states perfectly, but detects bright states as dark or neglects them. Therefore, the error of the double-threshold method is equal to $1/2$ for small measurement times. For long measurement times, the photon distribution for  initially dark and bright ions are nearly the same. If we choose $n_D$ and $n_B$ such that we neglect most of the overlap of both distributions, nearly no data will be left. The $\pi$-pulse method can show two different behaviors for long measurement times: (i) if we choose $n_c$ very large or very small, there will be also nearly no date left, (ii) if we choose $n_c$ in the middle, we will get an error of around $50\%$.

\section{Conclusion}

We generalize two detection methods \cite{Myerson2008,Hemmerling2012} for qubits with only a single possible state change during the detection process.  This generalized treatment is applicable to qubits  that undergo several state changes during the detection procedure such as, for example,  hyperfine qubits realized with trapped ions or neutral atoms, or solid state qubits such as NV centers in diamond. By introducing matrices of probabilities instead of single probability functions, numerical simulations as well as real-time experimental detection procedures of the generalized qubit detection methods can be efficiently implemented. Experiments carried out using a hyperfine qubit in  $^{171}Yb^+$ agree well with results of numerical simulations. 
Our results show that the generalized methods lead to smaller errors compared to the threshold method as well as compared to the original time-resolved method.

Furthermore, we introduce the double-threshold method. This method is a post-selective method similar to the $\pi$-pulse method. This method is applicable to qubits that undergo one ore more state changes, and also to qubits where (nearly) degenerate states are populated during a measurement. This is the case, for example, for hyperfine qubits. It ignores some data, however shows a decrease of the detection error.

Whereas we discuss in this paper the difference between one single possible state change and several possible state changes, there exists 
another difference between the original time-resolved and $\pi$-pulse method and the generalized methods: the rate of state change may not be given by nature (spontaneous decay), but can depend on experimental parameters such as the intensity of the laser inducing resonance fluorescence. Therefore, maximizing the fluorescence rate might not result in the minimum error  \cite{Ejtemaee2010}.  Future work will have to concerned with optimizing the generalized detection schemes taking explicitly into account adjustable experimental parameters.


\section*{Acknowledgment}

S.W. thanks O. G\"uhne for fruitful discussion. We acknowledge funding from Deutsche Forschungsgemeinschaft and from the European Community's Seventh
Framework Programme (FP7/2007-2013) under Grant Agreement No. 270843 (iQIT). 

\appendix

\section{Fluorescence rate of $^{171}Yb^+$\label{RB}}

\begin{figure}
\begin{center}
\includegraphics[width=0.4\textwidth]{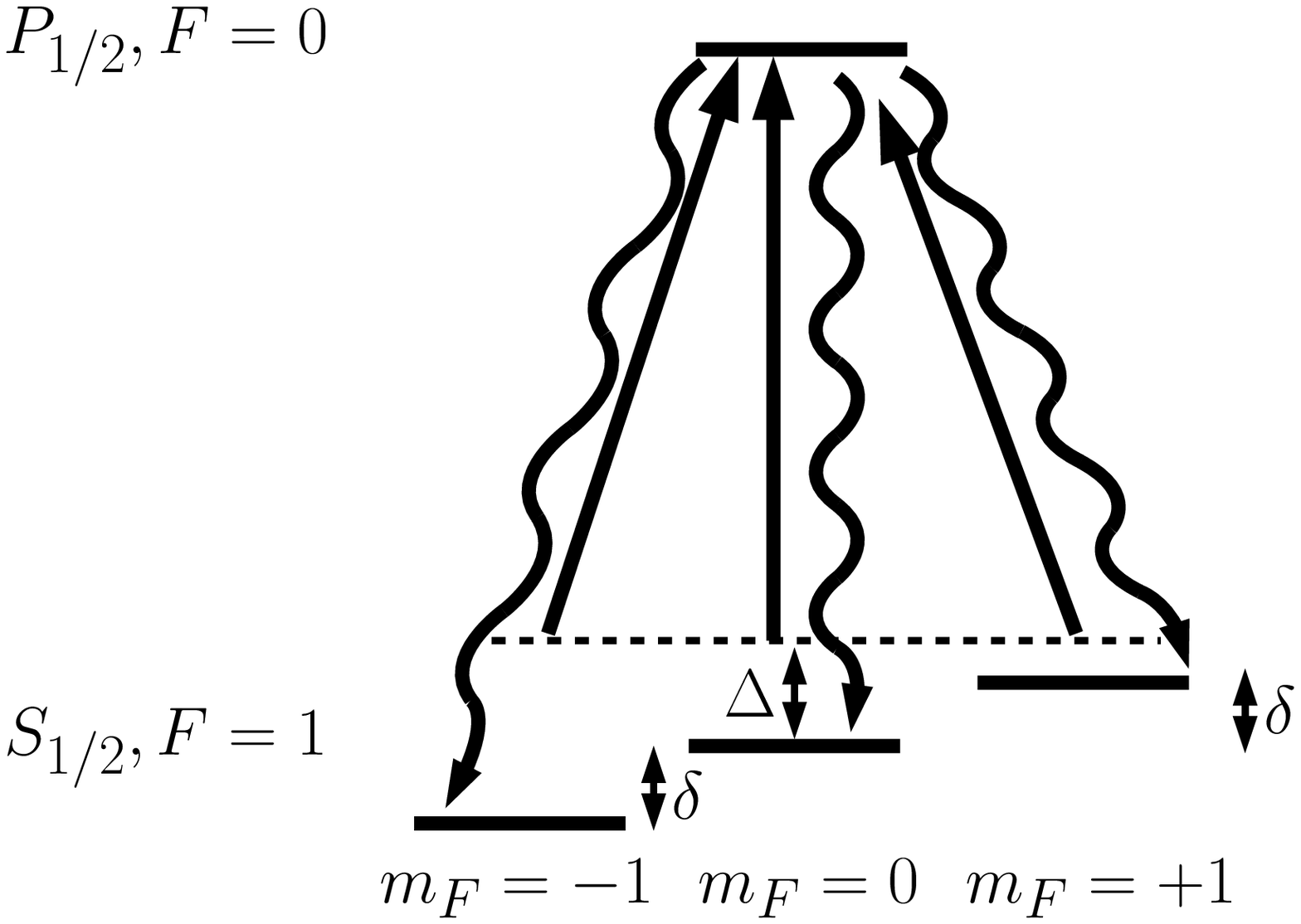}
\end{center}
\caption{Zeeman structure of the $S_{1/2},F=1 \leftrightarrow P_{1/2},F=0$ transition: the splitting due to the magnetic field is given by $\pm \delta$, the laser detuning $\Delta$ is defined relative to the magnetic field $B=0$. The branching for all decays is equal $1/3$   }
\label{aufspaltung}
\end{figure}

The fluorescence rate $R_B$ is given by 
\BE
R_B=\eta \cdot \gamma \cdot p_{f},
\EE
where $\eta$ is the photon collection efficiency, $\gamma$ the natural linewidth of the $P_{1/2}$ state and $p_f$ the steady state population of the $P_{1/2},F=0$ state. To calculate $p_f$ we have to include the Zeemann splitting of the $S_{1/2},F=0$ state (see \fig{aufspaltung}). To avoid dark states and to maximize $p_f$ a magnetic field has to be present and the laser needs to drive all transitions equally strong. For this case, $p_f$ is given by \cite{Ejtemaee2010}.
\BE
p_f=\frac{1}{36}\frac{\Omega^2}{\Delta^2+(\gamma'/2)^2}
\EE
with $\Omega$ the laser rabifrequency, $\Delta$ the detuning of the laser and
\BE
\left(\frac{\gamma'}{2}\right)^2=\left(\frac{\gamma}{2}\right)^2+\frac{1}{6}\left(\frac{\Omega^2}{36\delta^2}+4\delta^2\right).
\EE 

The photon rate $R_D$ is mainly given by light scattered by the ion or the apparatus and by blackcounts.

\section{Time-dependent mean photon rate\label{appendix}}

Phenomenologically we described the state changes of the ion by the probability
\BE
W_{BB}(t)=\E^{-t/\tau_B}
\EE
that a bright ion stays bright, and the probability
\BE
W_{DD}(t)=\E^{-t/\tau_D}
\EE
that a dark ion stays dark.  As a consequence, the state population $W_B$ ($W_D$) of the bright (dark) state are determined by the differential equations
\begin{eqnarray}
\dot{W}_B(t)&=& -\frac{1}{\tau_B}W_B(t)+\frac{1}{\tau_D}W_D(t)\\
\dot{W}_D(t)&=& \frac{1}{\tau_B}W_B(t)-\frac{1}{\tau_D}W_D(t).
\end{eqnarray}

For an initially bright ion we find the solution
\begin{eqnarray}
W_B^{(iB)}(t)&=&B+A\cdot \E^{-t/\tau}\\
W_D^{(iB)}(t)&=&A-A\cdot \E^{-t/\tau}
\end{eqnarray}
with $A=\tau_D/(\tau_B+\tau_D)$, $B=1-A$ and $\tau=\tau_B\tau_D/(\tau_D+\tau_B)=A\tau_B=B\tau_D$. The solution of an initially dark ion is given by
\begin{eqnarray}
W_B^{(iD)}(t)&=&B-B\cdot \E^{-t/\tau}\\
W_D^{(iD)}(t)&=&A+B\cdot \E^{-t/\tau}
\end{eqnarray}
In contrast to the probability distributions $W_{BB}$ and $W_{DD}$ the state populations $W_B^{(iB)}(t)$ and $W_D^{(iB)}(t)$ can be directly observed experimentally. As a consequence, the photon rate $r_B$  is given by
\BE
r_B(t)=R_B W_B^{(iB)}(t)+R_D
\EE
for an initially bright ion, and by
\BE
r_D(t)=R_B W_B^{(iD)}(t)+R_D
\EE
for an initially dark ion, with the fluorescence rate $R_B$ and the background photon rate $R_D$. The average photon number of an initially bright ion in the time interval $t_0-\Delta t\leq t\leq t_0$ is therefore equal to
\begin{eqnarray}
\overline{n}_B(t_0)&=&\int\limits_{t_0-\Delta t}^{t_0}r_B(t)\D t\nonumber\\&=&\Delta t (R_B B+R_D)+R_B A \tau (\E^{\Delta t/\tau}-1)\E^{-t_0/\tau}\nonumber \\
&=&a+b\E^{-t_0/\tau}.
\end{eqnarray}
Similar, we get for an initially dark ion
\begin{eqnarray}
\overline{n}_D(t_0)&=&\int\limits_{t_0-\Delta t}^{t_0}r_D(t)\D t\nonumber\\&=&\Delta t (R_B B+R_D)-R_B B \tau (\E^{\Delta t/\tau}-1)\E^{-t_0/\tau}\nonumber \\
&=&a-c\E^{-t_0/\tau}.
\end{eqnarray}
As a consequence, we get $A/B=b/c$ and with $B=1-A$ the result $A=(b/c)/(1+b/c)$ used to determined A with the help of the fit parameter $b$ and $c$. 

\section*{References}


\begin{thebibliography}{10}

\bibitem{Myerson2008}
A.~H. Myerson, D.~J. Szwer, S.~C. Webster, D.~T.~C. Allcock, M.~J. Curtis,
  G.~Imreh, J.~A. Sherman, D.~N. Stacey, A.~M. Steane, and D.~M. Lucas.
\newblock {\em Phys. Rev. Lett.}, 100:200502, 2008.

\bibitem{Hemmerling2012}
B.~Hemmerling, F.~Gebert, Y.~Wan, and P.~O. Schmidt.
\newblock {\em New J. of Phys.}, 14:023043, 2012.

\bibitem{Dehmelt1986}
W.~Nagourney, J.~Sandberg, and H.~Dehmelt.
\newblock {\em Phys. Rev. Lett.}, 56:2797, 1986.

\bibitem{Sauter1986}
Th~Sauter, W.~Neuhauser, R.~Blatt, and P.~E. Toschek.
\newblock {\em Phys. Rev. Lett.}, 57(14):1696--1698, 1986.

\bibitem{Bergquist1986}
J.~C. Bergquist, Randall~G. Hulet, Wayne~M. Itano, and D.~J. Wineland.
\newblock {\em Phys. Rev. Lett.}, 57(14):1699--1702, 1986.

\bibitem{Burrell2010b}
A.~H. Burrel, D.~J. Szwer, S.~C. Webster, and D.~M. Lucas.
\newblock {\em Phys. Rev. A}, 81:040302, 2010.

\bibitem{Wineland2011}
K.~R. Brown, A.~C. Wilson, Y.~Colombe, C.~Ospelkaus, A~M. Meier, E.~Knill,
  D.~Leibfried, and D.~J. Wineland.
\newblock {\em Phys. Rev. A}, 84:030303, 2011.

\bibitem{Blatt2009}
G.~Kirchmair, J.~Benhelm, F.~Z\"ahringer, R.~Gerritsma, C.~F. Roos, and
  R.~Blatt.
\newblock {\em Phys. Rev. A}, 79:020304, 2009.

\bibitem{Allcock2013}
D.~T.~C. Allcock, T.~P. Harty, C.~J. Ballance, B.~C. Keitch, N.~M. Linke, D.~N.
  Stacey, and D.~M. Lucas.
\newblock {\em Appl. Phys. Lett.}, 102:044103, 2013.

\bibitem{Dietrich2010}
M.~R. Dietrich, N.~Kurz, T.~Noel, G.~Shu, and B.~B. Blinov.
\newblock {\em Phys. Rev. A}, 81:052328, 2010.

\bibitem{Balzer2006}
Ch. Balzer, A.~Braun, T.~Hannemann, Ch. Paape, M.~Ettler, W.~Neunhauser, and
  Ch. Wunderlich.
\newblock {\em Phys. Rev. A}, 73:041407, 2006.

\bibitem{Olmschenk2007}
S.~Olmschenk, K.~C. Younge, D.~L. Moehring, D.~N. Matsukevich, P.~Maunz, and
  C.~Monroe.
\newblock {\em Phys. Rev. A}, 76:052314, 2007.

\bibitem{Ejtemaee2010}
S.~Ejtemaee, R.~Thomas, and P.~C. Haljan.
\newblock {\em Phys. Rev. A}, 82:063419, 2010.

\bibitem{Meyer2012}
H.~M. Meyer, M.~Steiner, L.~Ratschbacher, Ch. Zipkes, and M.~K\"ohl.
\newblock {\em Phys. Rev. A}, 85:012502, 2012.

\bibitem{Hensinger2013}
S.~Webster, S.~Weidt, J.~J. McLoughlin, and W.~K. Hensinger.
\newblock {\em Phys. Rev. Lett.}, 111:140501, 2013.

\bibitem{Mount2013}
E. Mount, S.-Y. Baek, M. Blain, D. Stick, D. Gaultney,
  S. Crain, R. Noek, T. Kim, P. Maunz, and J.g Kim.
\newblock {\em New J. of Phys.}, 15(9):093018, 2013.

\bibitem{Molmer2013}
S.~Gammelmark, W.~Alt, T.~Kampschulte, D.~Meschede, and K~M$\o$lmer.
\newblock arXiv 1312.5827.

\bibitem{Burrell2010}
A.~H. Burrell.
\newblock {\em High Fidelity Readout of Trapped Ion Qubits}.
\newblock PhD thesis, Exeter College, Oxford, 2010.

\bibitem{Khromova2012}
A.~Khromova, Ch. Piltz, B.~Scharfenberger, T.~F. Gloger, M.~Johanning, A.~F.
  Varón, and Ch. Wunderlich.
\newblock {\em Phys. Rev. Lett.}, 108:220502, 2012.

\bibitem{RAP}
Ch. Wunderlich, Th. Hannemann, T.~K\"orber, H.~H\"affner, Ch. Roos,
  W.~H\"ansel, R.~Blatt, and F.~Schmidt-Kaler.
\newblock {\em J. Mod. Opt.}, 54:1541, 2007.

\bibitem{Noek2013}
R. Noek, G. Vrijsen, D. Gaultney, E. Mount, T. Kim, P.
  Maunz, and J. Kim.
\newblock {\em Opt. Lett.}, 38(22):4735, 2013.

\end{thebibliography}

\end{document}